# Secure Geographic Routing Protocols: Issues and Approaches

[1]Mehdi sookhak, [2]Ramin Karimi, [4]Norafida Ithnin, [3]Mahboobeh Haghparast, [5]Ismail Fauzi ISnin

[1,2,3,5] Faculty of Computer Science and Information system
University technology Malaysia
Johor, 81300, Malaysia
(sookhak.mehdi, rakarimi1, nithnin) @gmail.com , ismailfauzi@utm.my

[4] Faculty of Library and Information Science
University of Isfahan, Iran
haghpanast@gmail.com

**Abstract**

In the years, routing protocols in wireless sensor networks (WSN) have been substantially investigated by researches. Most state-of-the-art surveys have focused on reviewing of wireless sensor network .In this paper we review the existing secure geographic routing protocols for wireless sensor network (WSN) and also provide a qualitative comparison of them.

## 1. Introduction

According to great capabilities of WSNs, application of them is increasing in recent decade. But, they face to some challenges such as limitation of power, memory, CPU and etc. these issues of WSNs have a direct effects on algorithms that are designed to them because complex algorithms need much memory and CPU and they consume a great deal of energy. These extreme limitations of resource, separate WSNs from traditional networks [1]. Based on the natural features of WSNs that distinguish them from other wireless networks such as ad hoc networks, routing in WSNs has very challenges. First, establishing comprehensive structure of address for deploying of the certain number of sensor nodes is impossible. So, traditional methods based on IP address (IP-based protocols) cannot be used to wireless sensor networks. Second, almost all applications of sensor networks need to sense the flow of data from multiple sources and transfer them to a special sink that it is as opposed to communication networks. Third if multiple sensors that are deployed in the adjacency of an event create same data, the data traffic is generated that it has an important redundancy in it. Such redundancy requires to be developed by the routing protocols to make energy and bandwidth utilization better. Finally, sensor node needs an accurate resource management because the resources of sensors such as energy, power of sending packet and the storage of sensor is restricted [2].

One of the important issues in WSNs is to provide the security of sensor nodes. There are various sensor holes as sink/black holes, worm holes, Sybil attack and etc. may form in a WSN and create network topology variations which trouble the upper layer applications [3]. Among all attacks, wormhole has more significant threat; because this type of attack does not need to compromise a sensor in the network and it can create the other type of attack easily. On the other hand, using a cryptographic technic cannot prevent wormhole attack [4].

## 2. Security Issues and Attacks on Sensor Network Routing

Most wireless sensor networks routing protocols are not complicated and they cannot protect themselves against large range of attacks. The attacks that can effect on WSNs are belonged to one of the following categorizations: spoofed, altered, or replayed routing information, selective forwarding, sinkhole attacks, Sybil attacks, wormholes, HELLO flood attacks, acknowledgement spoofing. The descriptions of each attack are mentioned in below [5].

### 2.1. Spoofed, Altered, or Replayed Routing Information

The main goal of most direct attack to routing protocol is to alter or modify the information that transmitted among nodes. create routing loops, attract or repel network traffic, extend or shorten source routes, generate false error messages, partition the network, increase end-to-end latency, etc. are some side effects of spoofing, altering, or replaying routing information on sensor networks.

## 2.2. Selective Forwarding attack

If an attacker intercepts or refuses to transmit a certain message and either drops it or chooses an arbitrary message for sending due to stop important message, the selective forwarding attack is occurred.

This attack may be appeared in two forms. In the simplest form of this attack, adversaries try to use a malicious node for rejecting and dropping all received packets. This type of selective forwarding attack operates like a black hole.

A second type of this attack happens when an adversary modifies transmitted packets. It is important to mention that the selective forwarding attack usually has most effect when the attacker is directly on the path of flowing data. But adversary can hear the neighbor packets from long distance [5].

## 2.3. Sinkhole attacks

The main goal of attacker in sinkhole is to attract large fractions of traffic to a region and constructing a sinkhole that the adversary is located in the center of it. For achieving this aim, Sinkhole attacks usually perform by making attractive a vulnerable node specifically to encircling nodes according to the routing algorithm. For example, an attacker can broadcast an advertisement or spoof for a very high quality route to a sink. Some protocols may try to confirm the truth of the quality of route with end-to-end acknowledgements including reliability or latency information.

The special communication pattern between sensors is one of the important reasons that sensor nodes are endangered from sinkhole attack. Since all packets in the network use and share only one base station, it is enough that compromised nodes find a single high quality route to the base station in order to influence a potentially large number of nodes [6,7].

## 2.4. Sybil attacks

The base of Sybil attack is that attacker can forge identities of nodes. a major side effect of this attack is to reduce the effectiveness of fault-tolerant schemes such as distributed storage, multipath routing, and topology maintenance.

Sybil attacks also pose a significant threat to geographic routing protocols. In geographical routing protocol, each node requires to transmit packet with its neighbors. So a node must have just a single set of coordinates from each of its neighbors and save them in its table but by utilizing the Sybil attack an attacker can be located in more than one situation at one time [5, 6].

## 2.5. Wormholes attack

In wormhole attack, attackers try to create a message appears that points away from the network. Wormhole attacks usually contain two malicious nodes that situated distant from each other. So, it can simply convince these two Separated nodes that they are neighbors by sending packets between the two of them. On the other hand, an adversary by using this attack could convince nodes that they are normally situated multiple hops from a base station that they are only one or two hops away. If an attacker is located near of sink or base station, it can interrupt routing by making a well-placed wormhole completely [6].

## 2.6. HELLO flood attack

Hello packets are a specific packet that usually used in many wireless sensor protocols. So, in these protocols each node needs to transmit HELLO packet for aware its neighbors, so that, when a node sends this packet, it may imagine that is located within radio range of the sender. Sometimes this assumption may be wrong. If an adversary transmits information with a sufficient power, every node in the network could be convinced that the attacker is its neighbor. This attack also can effect on protocols that based on localized information exchange between adjacent nodes like geographical routing protocol.

It is not essential for attackers to build lawful traffic due to utilize the HELLO flood attack. They can easily retransmit powerful overhead packets that every node in the network can received them [6].

## 2.7. Acknowledgment spoofing attack

The acknowledgment spoofing attack is designed based on this goal that a sender believes a frail connection is strong or that an unusable node is working. While nodes broadcast packet from weak or dead link, the packet may be lost. So, an attacker can prepare a selective forwarding attack utilizing acknowledgement spoofing by inspiring the certain node to send packets on those links [6,7].

## 3. Trust Issues

Trust and security are two important concepts that they are tightly interdependent. For example, cryptography is a modern technique for secure system that is dependent directly to a trusted key. One of the first definitions for trusted is based on Mayer, Davis and Schoorman (1995) "*the willingness of a party to be vulnerable to the actions of another party based on the expectation that the other party will perform a particular action important to the trust or, irrespective of the ability to monitor or control the party*" [8]. In wired networks, Trust is usually provided by applying indirect trust mechanisms, such as trusted certification agencies and authentication servers. But Trust establishment in wireless sensor networks is still an open and challenging field, because these trust relationships in such a networks are extremely susceptible to attacks. Also, the absence of fixed trust infrastructure, limited resources, ephemeral connectivity and availability, shared wireless medium and physical vulnerability, make trust establishment virtually impossible. To overcome these problems, to establish trust in wireless networks should be used a number of assumptions including pre-configuration of nodes with secret keys, or presence of an omnipresent central trust authority.

Asad Amir Pirzada and et al (2004) suggested and implemented a trusted model based on an effort/return mechanism. In this model, the trust is computed based on the information that each one node can gather from the other nodes in passive mode. By analyzing the received, forwarded and overheard packets, vital information about other nodes can be collected. Possible events that can be recorded in passive mode are included Frames received, Data packets forwarded, Control packets forwarded, Data packets received, Data forwarded, Data received, etc. Information that is retrieved from these events can be grouped into one trusted category and used to compute trust in other nodes in specific situations [9].

## 4. Overview of protocols

According to the large number of nodes that is deployed in the many of applications of sensor networks, it is impossible to dedicate comprehensive identifiers to each node. So, it is difficult to find out the unique way among sensors that deployed randomly for transmitting data. On the other hand, non-use of specific algorithms is not definitely useful regarding energy efficiency. Routing protocols is the best method to select a group of sensor nodes and applying data collection throughout the retransmission of data has been considered [10]. One of the important routing protocols in wireless sensor networks is Geographical routing protocol. The main strategy that used in geographical routing is named greedy forwarding in which the sender transmit packet to its neighbor that is located closest to the destination. There are several ideas to define the means of nearest node to destination such as Euclidean distance to the destination, the deviation from the imaginary line between source to destination and etc. [10,11].

In Following, some geographical routing protocols are reviewed briefly.

### 4.1. GPSR- Greedy Perimeter Stateless Routings

The Greedy Perimeter Stateless Routing is one of the usually used location-based routing protocols for launching and maintaining a sensor network. This protocol practically functions in a stateless manner and has the capability for multi-path routing. In GPSR, it is supposed that all nodes identify the geographical position of destination node with which communication is wanted. This location information (i.e.) geographical position is also used to route traffic to its required destination from the source node through the shortest path. Each transmitted data packet from node consist the destination node's identification and its geographical position similar two four-byte float numbers. Each node also frequently transmits a beacon to notify its near nodes relating to its recent geographical co-ordinates. The node positions are recorded, maintained and updated in a neighborhood table by all nodes receiving the beacon. To eliminate the overhead due to regular beacons, the node positions are carried onto forwarded data packets. GPSR supports two mechanisms for forwarding data packets: greedy forwarding and perimeter forwarding [12].

### 4.2. RGR-Receiver Based Forwarding for Geographical Routing Protocol.

Receiver Based Forwarding is an efficient approach for improving geographical sensor that is suggested and developed in December 2004 by Rodrigo Fonseca and et al. in Berkeley University. It is clear that in wireless sensor networks, when a message is transmit from one node to another; all the sender's neighbours can hear that message. According this feature, the main difference of this idea with GPSR is in packet forwarding because instead of sender decides to forward packet, the receivers determine next hop of packet. Scilicet, when a sender wants to transmit messages, instead of addressed to a specific neighbour, the receivers recognize that whether they should forward a message or not. As mentioned earlier, the flooding issue occurs when one node receives data packet, spreads it to all its neighbours. To prevent a flooding issue in this protocol, just if the location of neighbour is closer to destination than the previous sender, the message should be transmitted again. The computation of distance between each node from destination is done by using its coordinates. Also, each message contains a header that some information likes the coordinates of last sender and final node. So, with comparing the distance of current node to destination and the distance of pervious node to destination can decision about closeness to goal.

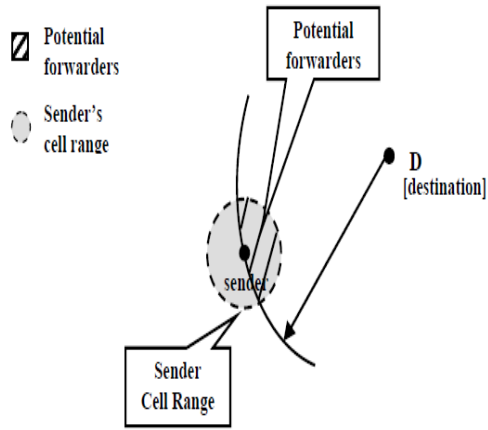

Figure1: Receiver-based protocol. Potential Forwarders (Fonseca & Sanz-Merino, 2004)

According this method, a neighbour that is nearer than sender can retransmit message but it is wasting energy. To avoid this approach, for all nodes was designed a timer which set it before forwarding message. Timer is adjusted based on closeness to destination. So, the neighbour that is closer to destination set a smaller timer than the nodes that are farther and its priority is higher. After that, if one neighbour listens to another neighbour forwarding while waiting for the timeout, it does not forward.

This routing protocol claim that it can prevent from spreading duplicates messages. So, if a node has broadcasted a copy of message, it does not retransmit following ones. To perform this, it is not sufficient to identify duplicates based on the header of message, because malicious nodes can modify the header of message. So, the identification method is carried out based on content and header of message [13].

### 4.3. S-GPSR–Secured Greedy Perimeter Stateless Routing Protocol.

As it mentioned before, GPSR is a routing protocol that used for geographic sensors but it is also exposed to various types of attackers. Another method which is suggested in 2010 by Samundiswary for protecting GPSR against some attacks such as sinkhole is called S-GPSR. In this method, is tried to joining trust based mechanism in the existing greedy perimeter stateless routing protocol prepares a secure routing protocol.

As mentioned in GPSR method, at first, each node during packet forwarding to a familiar destination must scan its neighbourhoods table to acquire the next hop which is optimal and leads to the goal. So, it selects the node that has the minimum distance to a specific destination. One of the newest methods for increasing the level of security in GPSR is using a trusted base approach in the neighbourhoods table to generate the most confident distance route rather than the default minimal distance. This is called S-GPSR.

The main component to implement the trust model in S-GPSR is Trust Update Interval (TUI) that used in each forward packet that is buffered in the nodes. The duration that each node should be waited before dedicating a trust or mistrust level to a node, is computed by TUI. Later than a node transmits a packet to its neighbour, it waits the neighbour's reaction for packet forwarding. So, this node faces to various situations. In the first case, the level trusted of node increased if neighbour forwards the packet in appropriate manner based on TUI. On the other hand, the level trust of node is declined if the packet is modified by the neighbour in an unsuitable way or it does not send the packet to next hop.

Each node in S-GPSR must perform two tasks, forward packet to its neighbour and control this packet. It is vital to check the integrity of forwarded packet by sender to verify the different fields in the forwarded IP packet. Therefore, confirming the acts of neighbour nods and enhancing the trust level is depended on succeeding the check of integrity. Vice versa, if the check of integrity fails or the neighbour node cannot broadcast packet, the node is treated as malicious node and the trust level decreases [14].

### 4.4. T-GPSR- Trusted Greedy Primeter Stateless Routing

During packet transmission to a known host, GPSR scans its neighbourhood table to retrieve the optimal next hop leading to the destination. As there may be more than one such hop available, GPSR selects an adjacent neighbour that has the least distance to a particular destination. In this protocol it is attempted to modify this rule and associate the computed trust level of a node along with its geographical position in the neighborhood table due to protect GPSR against Black hole attack. In order to create the most trusted route rather than the default minimal distance route, the trust levels are utilized with the geographical distances.

To implement the trust derivation mechanism, a node buffers (GPSR Agent::buffer packet) each forwarded packet for the Trust Update Interval (TUI). The TUI is a very critical component of such a trust model and determines the time a node should wait before assigning a trust or distrust level to a no debased upon the results of a particular event. After transmission, each node promiscuously listens for the neighbouring node to forward the packet. If the neighbour forwards the packet in the proper manner (correct modification if required) within the TUI, its corresponding trust level is incremented. However, if the neighbouring node modifies the packet in a unexpected manner or does not forward the packet at all, its trust level is decremented [15].

### 4.5. BSR-RRS and BSR-ANS –Boundary State Routing.

Contention BSR is implemented using the combination of Greedy Bounded Compass forwarding and the Boundary Mapping. In BSR protocol, Failure of geographic forwarding due to local minima only arises on void boundaries and the outer boundary. Previous research by Karp investigated the probing of boundaries to accumulate the link state information in boundary nodes. Boundary State Routing (BSR) relies upon Greedy-Bounded-Compass forwarding. Compass forwarding selects the neighbour on the closest angle to the destination. This protocol like GPSR does not have any security feature. In order to prevent wormhole attack against BSR, two methods is designed that called BSR-RRS and BSR-ANS. Reverse Route Scheme (RRS)use hop-count technic to find malicious nodes but Authentication of Nodes Scheme (ANS) is based on authentication to find the not honest nodes [16].

## 5. Simulation and analysis

Simulation is one of the important steps in any survey because it allows to investigator for simulating and testing its idea in the virtual area that likes a real world. In order to simulate these routing protocols, NS-2 is selected. It is assumed that among 50 to 200 nodes are deployed randomly in 500*500 areas. In the following table some of the important parameters are mentioned.

| Simulation Parameters | Values |
|---|---|
| Number of Nodes | 50 and 200 |
| Geographical environment | 500*500 |
| Size of Packet (bytes) | 512 |
| Traffic Type | CBR |
| Number of malicious nodes | Depend on type of attack (2-25) |
| Mobility model | Depend on type of routing (Static or Random way point) |
| Pause time(s) | 20 |
| Simulation time(s) | 100 |

Table 1: simulation parameters

Deliver ratio is one of the useful measurement parameters in order to prove the efficiency of these secure protocol in which it is tried to calculate the numbers of packets received by destination nodes divide to the number of packets are sent by source nodes.

### 5.1. T-GPSR and GPSR against Blackhole Attack

As it is mentioned, T-GPSR is a protocol that is designed to protect GPSR against Black hole attack. In order to check the efficiency of this protocol, 50 nodes are deployed in the area randomly. This protocol support random mobility in any way. As it is shown in the following chart, the T-GPSR has a better reaction against this attack when the numbers of malicious nodes increase. This method can improve the delivery ratio rate to 80 percent.

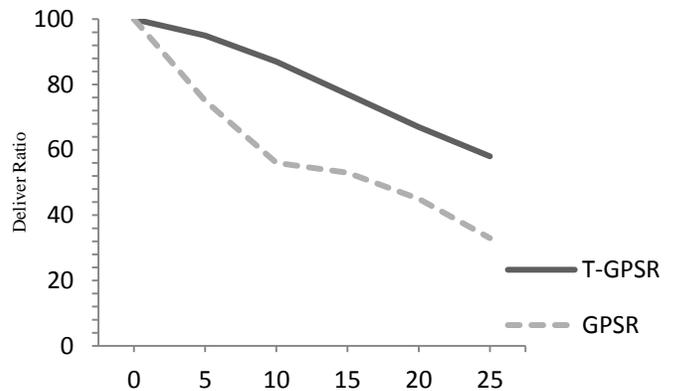

Figure 2: Deliver ratio for T-GPSR and GPSR against Black hole Attack

### 5.2. S-GPSR and GPSR against Selective forwarding Attack

The main purpose to design S-GPSR is to protect GPSR against selective forwarding attack by using trusted model. To evaluate this method, 100 nodes are deployed in the 500*500 (m2) environments. When the number of malicious nodes is among 5 to 15, deliver ratio of S-GPSR is about 70 percent. Finally, the rate of delivery in S-GPSR is more than GPSR, clearly.

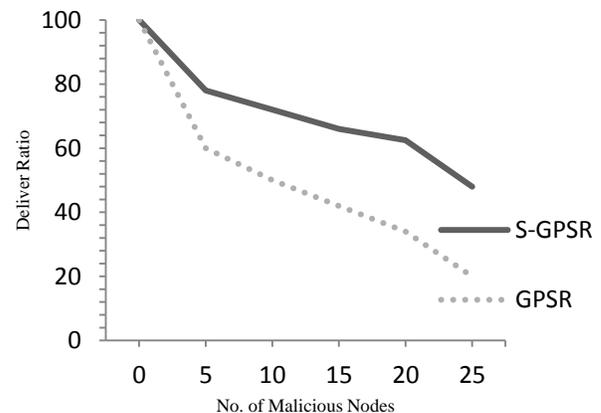

Figure3: Deliver ratio for S-GPSR and GPSR against Selective Forwarding Attack (Mobile)

### 5.3. RGR and GPSR against Selective Forwarding and Wormhole Attack

RGR is a method that was suggested for static geographic routing protocol based on GPSR in order to increase the security level of it. The rate of packet delivery for RGR and GPSR are illustrated in figure 4. This chart is shown thatRGR by using multipath method is more protected against selective forwarding attack than GPSR. For example, when 10 malicious nodes exist in the network, RGR's delivery ratio is higher than 80 percent, approximately. (It is important to mention that RGR is usable for static sensor networks but S-GPSR support mobility, so the comparison of them is not logical.)

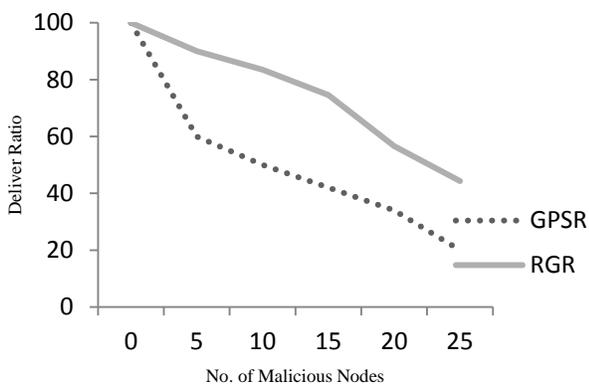

Figure 4: Deliver ratio for RGR and GPSR against Selective Forwarding Attack (Static)

In the next situation, it is tried to prove that RGR is protected against Wormhole attack but GPSR do not have any features against this attack. As it is shown in the following figure, the rate of delivery is approached to less than 10 percentwhen the number of malicious nodes is more than 4 in GPSR but RGR has a better reaction against this attack.

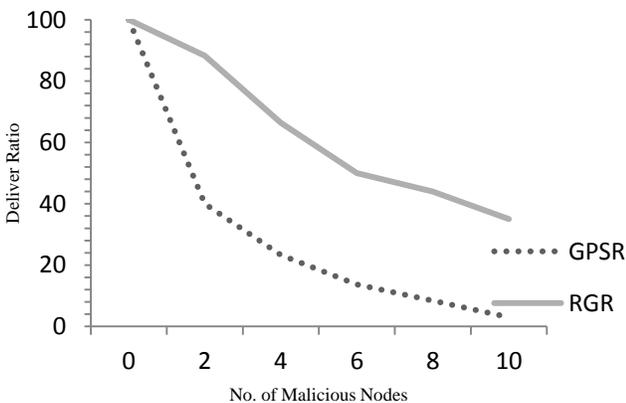

Figure 5: Deliver ratio for RGR and GPSR against Wormhole Attack (Static)

### 5.4. BSR, BSR-RRS andBSR-ANS against Wormhole Attack

BSR is a geographic routing protocol that used Greedy-Bounded-Compass to forward packet through destination in which there are not any security features. To secure this method against wormhole attack two method was suggested. BSR-RRS is the first method that tries to identify wormhole attack by utilizing Hop-Count technic. Based on this model, the number of hop from source to destination is compared to the number of hops through destination to source. In the next method that is called BSR-ANS, use cryptographic authentication in order to find the malicious node. The following figure is shown the rate of packet that is received in the destination to the number of packets which are sent. It is clear that ANS is the best model among these three methods by using digital signature of nodes. As it clear RRS cannot protect sensor network against wormhole attack completely.

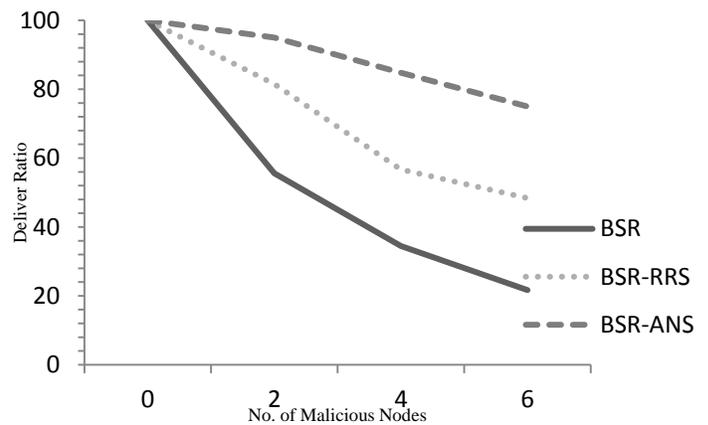

Figure 6: Deliver ratio in BSR-ANS is more than BSR-RRS and BSR against wormhole attack with mobility

### 6. Conclusion

In this paper, we review the secure geographic routing protocols. We also discuss about why some of routing protocols protect against some attack. Hence there are metrics to evaluate the protocols namely localization information (GPS), authentication, integrity and trust modeIn order to improve their level of security. We simulated the protocols based on the delivery ratio. A qualitative comparison of secure routing protocols is summarized in table 2.

| Routing Protocol | Localization Information(GPS) | Authentication | Integrity | Trust model |
|---|---|---|---|---|
| TGPSR | ✓ | No | ✓ | ✓ |
| SGPSR | ✓ | No | ✓ | ✓ |
| RGR | ✓ | No | ✓ | No |
| BSR-RRS | ✓ | No | No | No |
| BSR-ANS | ✓ | ✓ | No | No |

TABLE 2: QUALITATIVE COMPARISON OF SECURE ROUTING PROTOCOLS

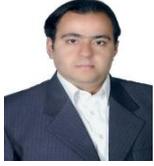
**Mehdi Sookhak** received the B.Sc. degree in computer engineering from Azad University, Shiraz, Iran, in 2001. Now, he is student of the M.A.Sc. degree in computer Science (information security) from the University Technology Malaysia. He has several publications in International Conferences and journal such as "3rd International Conferences of Engineering, Science and Humanities" in Malaysia, "3rd International Conference on Computer Technology and Development", Chengdu, China, 2011 and International Journal of Computer Science Issues, July 2011. His current research interests include security of wireless sensor networks.

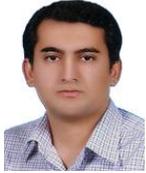
**Ramin Karimi** is currently a Ph.D candidate in Department of Computer Science and Information Technology at Universiti Teknologi Malaysia, Johor, Malaysia. He received M.Sc degree in computer engineering from Iran University of Science and Technology in 2006. His research interests include Vehicular Ad Hoc Networks, Mobile ad-hoc networks and communication Networks.

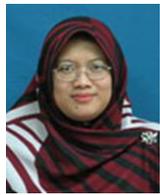
**Norafida Ithnin** is a senior lecturer at Universiti Teknologi Malaysia. She received her B.Sc degree in computer science from Universiti Teknologi Malaysia in 1995, her MSc degree in Information Teknologi from University Kebangsaan Malaysia in 1998 and her PHD degree in computer science from UMIST, Manchester in 2004. Her primary research interests are in security, networks, Mobile ad-hoc networks, Vehicular Ad Hoc Networks.

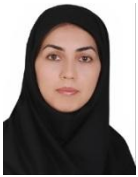
Mahboobeh Haghparast received M.A. degree in library and information science from Isfahan University of Iran in2010. Her research interest includes information science and information systems.